\documentclass[preprint,12pt]{elsarticle}

\usepackage[utf8]{inputenc}
\usepackage{amssymb}
\usepackage[svgnames,dvipsnames,x11names]{xcolor}
\usepackage{hyperref}
\usepackage{amsmath,amssymb,amsbsy,amstext,amsfonts,bm}

\journal{Nuclear Physics B}

\begin{document}

\begin{frontmatter}



\title{Primordial black holes and secondary gravitational waves from natural inflation}


\author[label1]{Qing Gao}
 \ead{gaoqing1024@swu.edu.cn}
 \affiliation[label1]{organization={School of Physical Science and Technology, Southwest University},
             city={Chongqing},
             postcode={400715},
             country={China}}

\author[label2]{Yungui Gong\corref{cor1}}
 \ead{yggong@hust.edu.cn}
 \affiliation[label2]{organization={School of Physics, Huazhong University of Science and Technology},
             city={Wuhan},
             postcode={430074},
             state={Hubei},
             country={China}}
\cortext[cor1]{Corresponding author}

\author[label3]{Zhu Yi}
 \ead{yz@bnu.edu.cn}
 \affiliation[label3]{organization={Department of Astronomy, Beijing Normal University},
             city={Beijing},
             postcode={100875},
             country={China}}



\begin{abstract}
The production of primordial black hole (PBH) dark matter (DM) and the
generation of scalar induced secondary gravitational waves by using
the enhancement mechanism with
a peak function in the non-canonical kinetic term in natural inflation is discussed.
We show explicitly that the power spectrum for the primordial curvature perturbation can be enhanced at $10^{12}$ Mpc$^{-1}$, $10^{8}$ Mpc$^{-1}$ and $10^{5}$ Mpc$^{-1}$
by adjusting the model parameters.
With the enhanced primordial curvature perturbations,
we show the production of PBH DM with peak masses around
$10^{-13}\ M_{\odot}$, the Earth's mass and the stellar mass,
and the generation of scalar induced gravitational waves (SIGWs) with peak frequencies around mHz, $10^{-6}$ Hz and nHz, respectively.
The PBHs with the mass scale $10^{-13}\ M_{\odot}$ can
make up almost all the DM and the associated SIGWs is testable by spaced based gravitational wave observatory.
\end{abstract}







\end{frontmatter}


\section{Introduction}	
\label{sec1}
If the density contrast of overdense regions at the horizon reentry exceeds the threshold value during radiation domination,
then the overdense regions gravitationally collapse to form primordial black holes (PBHs) \cite{Carr:1974nx,Hawking:1971ei}.
Gravitational waves (GWs) detected by the Laser Interferometer Gravitational Wave Observatory (LIGO) Scientific Collaboration and the Virgo Collaboration
\cite{Abbott:2016blz,Abbott:2016nmj,Abbott:2017vtc,Abbott:2017oio,TheLIGOScientific:2017qsa,Abbott:2017gyy,LIGOScientific:2018mvr,Abbott:2020uma,LIGOScientific:2020stg,Abbott:2020khf,Abbott:2020tfl,Abbott:2020niy}
may be emitted by PBHs \cite{Bird:2016dcv,Sasaki:2016jop}.
PBHs may account for dark matter (DM)
\cite{Ivanov:1994pa,Frampton:2010sw,Belotsky:2014kca,Khlopov:2004sc,Clesse:2015wea,Carr:2016drx,Inomata:2017okj,Garcia-Bellido:2017fdg,Kovetz:2017rvv,Carr:2020xqk}
and even explain the planet 9 \cite{Scholtz:2019csj}.
Because of these reasons, the recent interest in the seeds of PBHs from primordial curvature  perturbations generated during inflation arises.

To produce abundant PBH DM, the amplitude of primordial curvature perturbations at small scales needs to be in the order of $0.01$ \cite{Lu:2019sti,Sato-Polito:2019hws} while at large scales the cosmic microwave background (CMB) constraint on the amplitude of the power spectrum is
$A_s=2.1\times 10^{-9}$ \cite{Akrami:2018odb}.
Inflationary models with inflection point or non-minimal coupling are usually
considered to enhance the power spectrum at small scales
\cite{Gong:2017qlj,Martin:2012pe,Motohashi:2014ppa,Garcia-Bellido:2017mdw,Ezquiaga:2017fvi,Germani:2017bcs,Motohashi:2017kbs,Bezrukov:2017dyv,Espinosa:2017sgp,Cicoli:2018asa,Sasaki:2018dmp,Kamenshchik:2018sig,Gao:2018pvq,Dalianis:2018frf,Ballesteros:2018wlw,Passaglia:2018ixg,Passaglia:2019ueo,Bhaumik:2019tvl,Dalianis:2019vit,Fu:2019ttf,Fu:2019vqc,Xu:2019bdp,Braglia:2020eai,Gundhi:2020zvb,Zhou:2020kkf}.
Usually it is very difficult to enhance the power spectrum
to the order of 0.01 at small scales while keeping the total number of e-folds to be 50-60 \cite{Sasaki:2018dmp,Passaglia:2018ixg}.
For canonical single field inflationary models with inflection point, the power spectrum either cannot be enhanced to the order of $0.01$ \cite{Garcia-Bellido:2017mdw,Ezquiaga:2017fvi,Germani:2017bcs,Passaglia:2018ixg} or the model parameters need to be fine tuned by more than six decimal digits \cite{Gong:2017qlj,Cicoli:2018asa,Gao:2018pvq,Passaglia:2018ixg}.
By introducing a non-canonical kinetic term like $k$ inflation \cite{ArmendarizPicon:1999rj,Garriga:1999vw}
and $G$ inflation \cite{Kobayashi:2010cm,Kobayashi:2011nu,Kobayashi:2011pc,Herrera:2018ker},
a new mechanism with a peak function in the non-canonical kinetic term
was proposed to enhance the primordial power spectrum at small scales \cite{Lin:2020goi,Yi:2020kmq,Yi:2020cut}.
The forms of the peak function and the inflationary potential in the mechanism
are not restricted, and both sharp and broad peaks are possible \cite{Yi:2020kmq,Yi:2020cut,Gao:2021vxb}.
The mechanism works for both Higgs inflation and T-model.
To the second order of perturbations,
the first order perturbations are the sources of the second order tensor perturbation, so associated with the formation of PBHs
the large curvature perturbations at small scales induce secondary GWs after the horizon reentry during  the radiation dominated epoch
\cite{Matarrese:1997ay,Mollerach:2003nq,Ananda:2006af,Baumann:2007zm,Garcia-Bellido:2017aan,Saito:2008jc,Saito:2009jt,Bugaev:2009zh,Bugaev:2010bb,Alabidi:2012ex,Orlofsky:2016vbd,Nakama:2016gzw,Inomata:2016rbd,Cheng:2018yyr,Cai:2018dig,Bartolo:2018rku,Bartolo:2018evs,Kohri:2018awv,Espinosa:2018eve,Cai:2019amo,Cai:2019elf,Cai:2019bmk,Cai:2020fnq,Domenech:2019quo,Domenech:2020kqm,Pi:2020otn,Fumagalli:2020adf,Fumagalli:2020nvq}.
These scalar induced GWs (SIGWs) consist  of  the  stochastic  background
and they can detected by Pulsar Timing Arrays (PTA) \cite{Ferdman:2010xq,Hobbs:2009yy,McLaughlin:2013ira,Hobbs:2013aka,Moore:2014lga}
and the space based GW observatories like Laser Interferometer Space Antenna (LISA) \cite{Danzmann:1997hm,Audley:2017drz}, Taiji \cite{Hu:2017mde} and TianQin  \cite{Luo:2015ght}.

Nambu-Goldstone bosons are ubiquitous because they arise whenever a global symmetry is spontaneously broken.
These particles become pseudo-Nambu-Goldstone bosons (PNGBs) if additional explicit symmetry is broken.
For axions, the mass of PNGBs arises from nonperturbative instantons through the chiral anomaly. Instanton effects produce a periodic potential of height $\Lambda^4$ for PNGBs when the associated gauge group
becomes strong at a mass scale $\Lambda$. Taking the symmetry breaking scale $f_a$ as the Planck scale $M_\text{pl}$ and
the mass scale $\Lambda$ as the scale of grand unification $10^{15}$ GeV,
we get a naturally flat potential without any fine tuning.
The inflationary model with this flat potential is called natural inflation \cite{Freese:1990rb}.

In this paper, we use the enhancement mechanism \cite{Yi:2020kmq,Yi:2020cut} to discuss the production of PBHs and SIGWs in natural inflation.
The paper is organized as follows.
In Sec. \ref{sec2}, we discuss the enhancement of the power spectrum, the production of PBHs and the generation of SIGWs in natural inflation.
The conclusions are drawn in Sec. \ref{sec3}.

\section{Natural Inflation}
\label{sec2}
In this section, we consider natural inflation with a non-canonical kinetic term.
The action is
\begin{equation}\label{act1}
  S=\int d x^4 \sqrt{-g}\left[\frac{1}{2}R+X+G(\phi)X-V(\phi)\right],
\end{equation}
where $X=-g_{\mu\nu}\nabla^{\mu}\phi\nabla^{\nu}\phi/2$,
the reduced Planck mass is taken to be $M^{-2}_\text{Pl}=8\pi G=1$,
and we introduce
noncanonical kinetic terms like those in scalar-tensor theory of gravity,
k inflation \cite{ArmendarizPicon:1999rj,Garriga:1999vw} or G inflation \cite{Kobayashi:2010cm}.
The potential for natural inflation is \cite{Freese:1990rb}
\begin{equation}\label{pngb}
 V(\phi)=\Lambda^4\left[1+\cos\left(\frac{\phi}{f_a}\right)\right],
\end{equation}
where $\Lambda\sim 10^{15}$ GeV is the energy scale of the potential,
and $f_a\sim M_\text{Pl}$ is the symmetry breaking scale.
The noncanonical kinetic function $G(\phi)=G_p(\phi)+f(\phi)$ \cite{Yi:2020kmq,Yi:2020cut} with the function
\begin{equation}
\label{fphieq}
f(\phi)=f_0\left[1+\cos(\phi/f_a)\right]^4\left[\frac{\sin(\phi/f_a)}{f_a}\right]^2,
\end{equation}
and the peak function $G_p(\phi)$.
Motivated by the noncanonical kinetic term $X/\phi$ in Brans-Dicke theory,
we can take $G_p(\phi)=hw/(\phi-\phi_p+w)$ with $h w\sim O(1)$ and the dimensionless parameter $w\ll 1$ to avoid singularity at $\phi=\phi_p$.
In this paper, we take more general peak function
\cite{Yi:2020kmq,Yi:2020cut}
\begin{equation}\label{gfuncng}
  G_p(\phi)=\frac{h}{1+\left(|\phi-\phi_p|/{w}\right)^q},
\end{equation}
where the dimensionless parameter $h$ determines the height of the peak,
the dimensionless parameter $w$ controls the width of the peak,
$\phi_p$ normalized by the reduced Planck mass $M_\text{Pl}$ controls the peak position of the enhanced power spectrum,  the power index $q$ controls the shape of the
enhanced power spectrum,
and $f_0$ has the dimension of mass.
Note that at low energies after inflation, the non-canonical kinetic function $G(\phi)$ is negligible and the standard canonical kinetic term is recovered.
The background equations are
\begin{equation}
\label{Eq:eom1}
3H^2=\frac{1}{2}\dot{\phi}^2+V(\phi)+\frac{1}{2}\dot{\phi}^2G(\phi),
\end{equation}
\begin{equation}
\label{Eq:eom2}
\dot{H}=-\frac{1}{2}[1+G(\phi)]\dot{\phi}^2,
\end{equation}
\begin{equation}
\label{Eq:eom3}
\ddot{\phi}+3H\dot{\phi}+\frac{V_{\phi}+\dot{\phi}^2G_{\phi}/2}{1+G(\phi)}=0,
\end{equation}
where $G_\phi=dG(\phi)/d\phi$, $V_\phi=dV/d\phi$.

\subsection{The enhancement of the scalar power spectrum}
To understand how the mechanism can enhance the power spectrum qualitatively,
first we use slow-roll formulae to explain it.
For the calculation of the power spectrum, we do not assume slow-roll and instead invoke numerical method to solve both the background and perturbation equations.
The equation for the curvature perturbation $\zeta$ is
\begin{equation}\label{zeta:k}
  \frac{d ^2 u_k}{d\eta^2}+\left(k^2-\frac{1}{z}\frac{d z^2}{d\eta^2}\right)u_k=0,
\end{equation}
where $\eta$ is the conformal time $\eta=\int dt/a(t)$, $u_k=z\zeta_k$,  $z=a\dot{\phi}(1+G)^{1/2}/H$.
In the slow-roll approximation,
the power spectrum for the curvature perturbation is
\begin{equation}
\label{ps}
\begin{split}
\mathcal{P}_\zeta=&\frac{k^3}{2\pi^2}\left|\zeta_k\right|^2\\
=&\frac{H^4}{4\pi^2\dot{\phi}^2(1+G)}
\approx \frac{V^3}{12\pi^2V_{\phi}^2}(1+G).
\end{split}
\end{equation}
The scalar spectral index $n_s=1+d\ln\mathcal{P}_\zeta/d\ln k$.
The tensor perturbations are not affected by the non-canonical kinetic term and the power spectrum for tensor perturbations is
\begin{equation}
\label{ts1}
\mathcal{P}_T\approx 6 H^2/(2\pi)^2.
\end{equation}
The tensor to scalar ratio is $r=\mathcal{P}_T/\mathcal{P}_\zeta$.

\begin{figure}[htbp]
 \centering
 \includegraphics[width=0.7\textwidth]{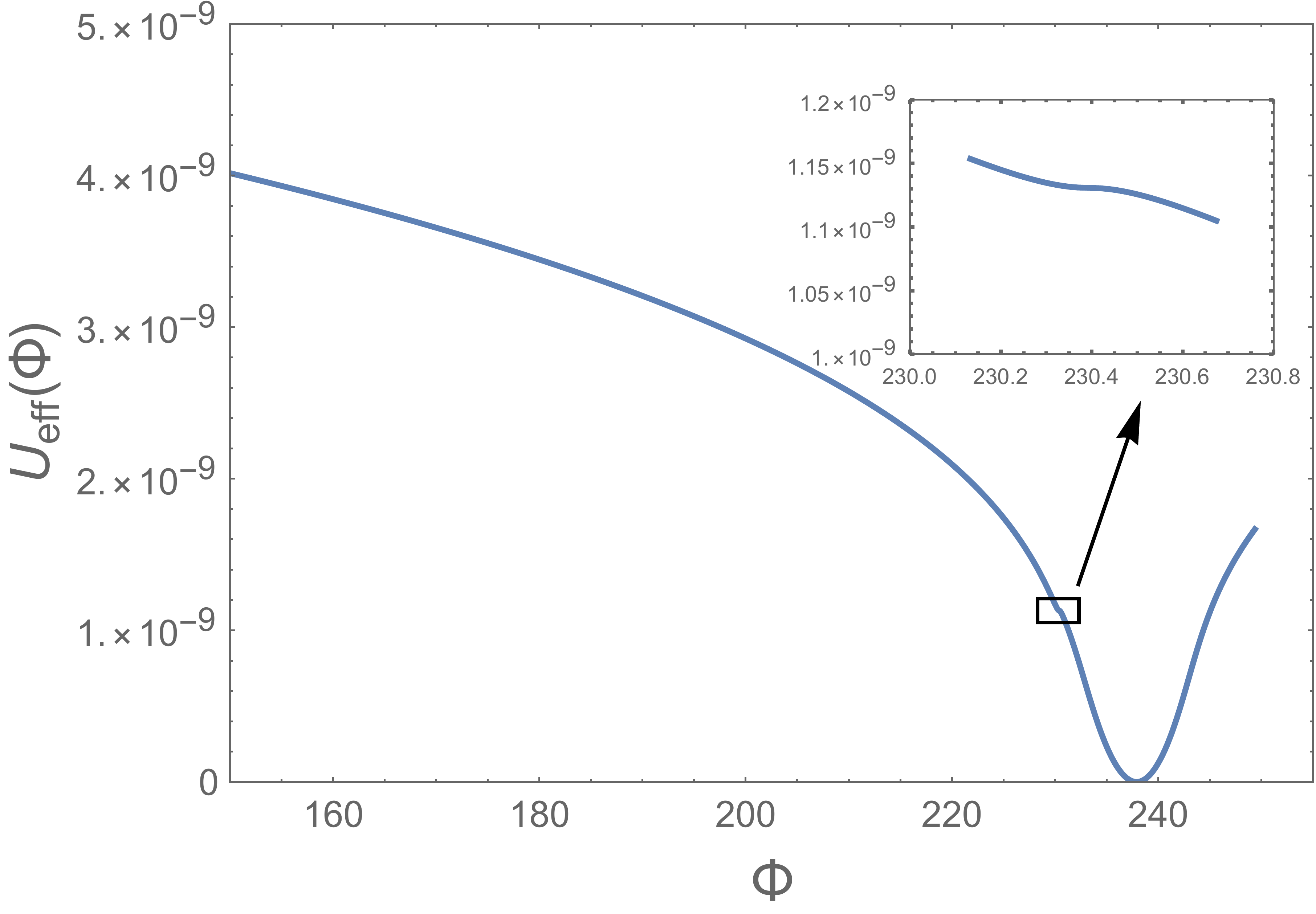}
 \caption{The effective potential of the canonical field $\Phi$ for the model N1. The inset shows the inflection point around $\phi_p$.}\label{ueff}
\end{figure}

From Eq. \eqref{ps}, we see that the non-canonical function $G(\phi)$
can enhance the scalar power spectrum.
Around the peak $\phi_p$, the major contribution to $G(\phi)$ is from $G_p(\phi)$ and $G(\phi)\approx h$, so the scalar power spectrum can be enhanced by $h$.
To achieve seven orders of magnitude enhancement,
$h$ should be at least the order of $10^7$.
On the other hand,
the number of $e$-folds  before the end of inflation when the pivotal scale exits the horizon is
\begin{equation}\label{efold0}
N=\int_{\phi_e}^{\phi_*}\frac{V}{V_{\phi}}d \phi+\frac{V(\phi_p)}{V_{\phi}(\phi_p)}\int_{\phi_p+\Delta \phi}^{\phi_p-\Delta \phi}G_p(\phi) d\phi,
\end{equation}
so the enhancement also increases $N$ and
the contribution from the peak function is about $20$ $e$-folds,
here $\phi_*$ and $\phi_e$ are the values of the scalar field
at the horizon exit and the end of inflation, respectively.
This means that when the peak function $G_p(\phi)$ enhances the scalar power spectrum at small scales,
it also effectively moves $\phi_*$ closer to $\phi_e$
and the remaining number of $e$-folds $N_{eff}\sim 40$ for the slow-roll inflation since the usual slow-roll inflation is recovered away from the peak.
This changes the predictions of
the scalar spectral tilt and the tensor to scalar ratio for natural inflation at large scales.
Away from the peak, the peak function is negligible and the non-canonical function $f(\phi)$ dominates,
we can take the transformation \cite{Yi:2020kmq,Yi:2020cut}
\begin{equation}\label{trans}
  d\Phi=\sqrt{1+f(\phi)}d\phi
\end{equation}
to change the non-canonical scalar field $\phi$ to be canonical scalar field $\Phi$
with the potential $U(\Phi)=V[\phi(\Phi)]=U_0\Phi^{1/3}$
and $U_0=(9\Lambda^{24}/f_0)^{1/6}$.
The expression \eqref{fphieq} for the function $f(\phi)$ is derived from the above transformation \eqref{trans}.
For the power law potential $U(\Phi)=U_0\Phi^{1/3}$ and $N_{eff}\sim 40$, we get
$n_s=0.971$ and $r=0.033$ which are consistent with Planck 2018 results. If we use the full peak function $G(\phi)$ in Eq. \eqref{trans} to change the non-canonical scalar field $\phi$ to be canonical scalar field $\Phi$,
then we can get the effective potential $U_{eff}(\Phi)$ of the canonical field $\Phi$.
In general, there is no analytical relation between $\phi$ and $\Phi$ and we cannot obtain the expression for the effective potential.
However, we can use numerical method to get the effective potential $U_{eff}(\Phi)$ as shown in Fig. \ref{ueff} for the model N1 defined below.
The effective potential $U_{eff}(\Phi)$ of the canonical field $\Phi$ has an inflection point.
The existence of inflection point explains the enhancement of the power spectrum.

\begin{figure}[htbp]
 \centering
 \includegraphics[width=0.7\textwidth]{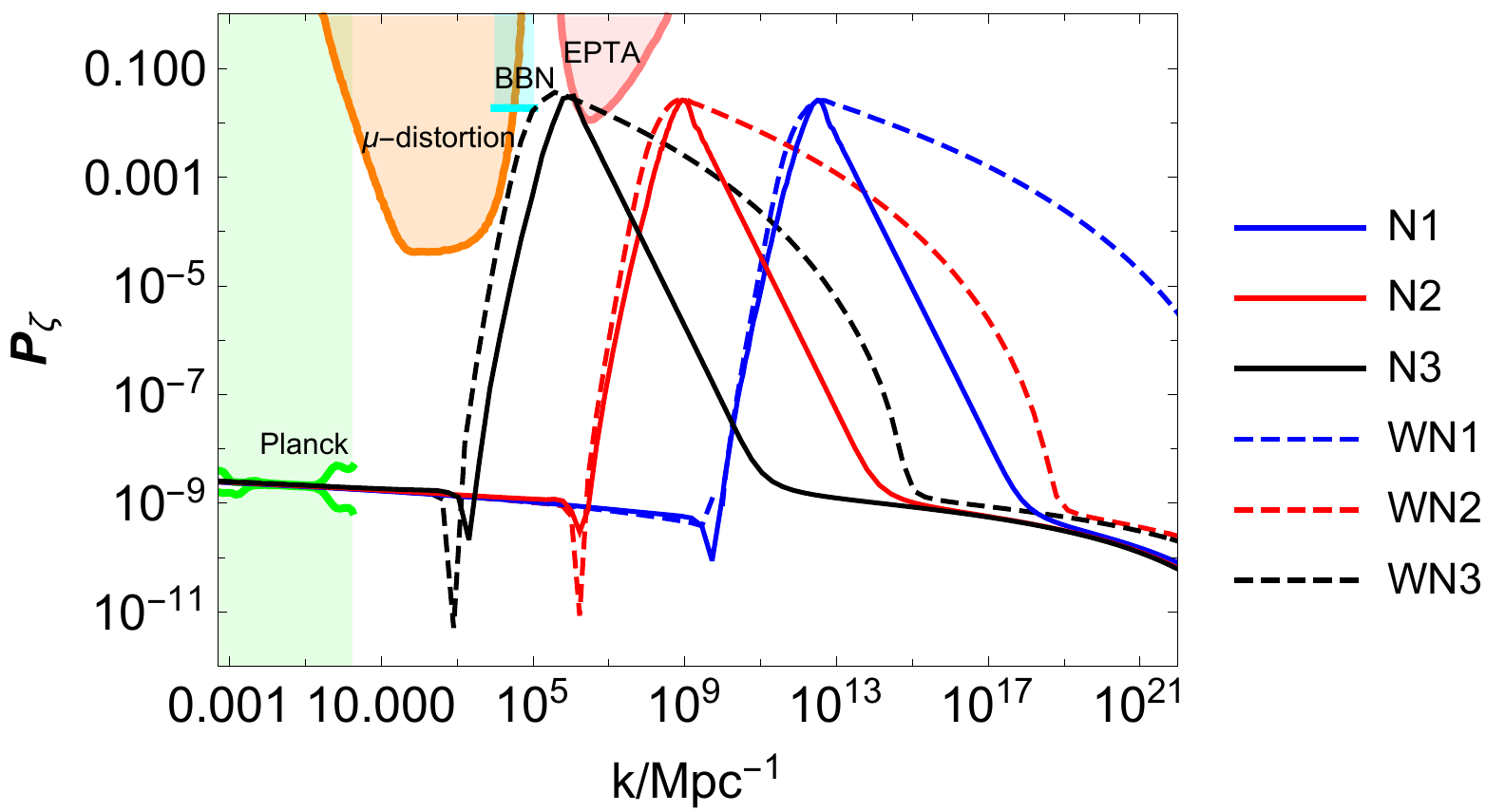}
 \caption{The results for the scalar power spectrum.
The solid lines denote the models with the parameter $q=1$
and the dashed lines denote the models with the parameter $q=6/5$.
The blue, red, black lines denote the models with the peaks around
$10^{12}$ Mpc$^{-1}$, $10^{8}$ Mpc$^{-1}$ and $10^{5}$ Mpc$^{-1}$, respectively.
The parameters for the models and the peak scales $k_{\text{peak}}$ are shown in Table \ref{table1}.
The peak values of the power spectra are shown in Table \ref{table2}.
The lightgreen shaded region is excluded by the CMB observations \cite{Akrami:2018odb}. The pink, cyan and orange regions show the
constraints from the PTA observations \cite{Inomata:2018epa},
the effect on the ratio between neutron and proton
during the big bang nucleosynthesis (BBN) \cite{Inomata:2016uip}
and $\mu$-distortion of CMB \cite{Fixsen:1996nj}, respectively.}\label{pr}
\end{figure}

Therefore, we see that for natural inflation it is possible to satisfy the large scale constraints and enhance the scalar power spectrum at small scales.
To show this, we numerically solve the background Eqs. \eqref{Eq:eom1}-\eqref{Eq:eom3}
and the perturbation Eq. \eqref{zeta:k} to get both the scalar and tensor power spectra $\mathcal{P}_\zeta$ and $\mathcal{P}_T$.
Then we numerically calculate the scalar spectral index $n_s$
and the tensor to scalar ratio $r$ from $\mathcal{P}_\zeta$ and $\mathcal{P}_T$.
We take $f_0=7500$, $\phi_*=14.5$, the symmetry breaking scale $f_a=7$,
and the values of the parameters $\Lambda$, $h$ and $\phi_p$ as shown in Table \ref{table1}.
In table \ref{table1}, we use the label "N" to represent the model with the parameter $q=1$ which produces sharp peaks at small scales in the scalar power spectrum,
and the label "WN" to represent the model with the parameter $q=6/5$ which produces broad peaks at small scales in the scalar power spectrum.
We also use the labels 1, 2, 3 to distinguish the models with different peak scales in the scalar power spectrum,
the peak scale for the model labelling as 1 is around $10^{12}$ Mpc$^{-1}$,
the peak scale for the model labelling as 2 is around $10^{8}$ Mpc$^{-1}$,
and the peak scale for the model labelling as 3 is around $10^{5}$ Mpc$^{-1}$.
The numerical results for $N$, $n_s=1+d\ln \mathcal{P}_\zeta/d\ln k$, the tensor to scalar ratio $r=\mathcal{P}_\zeta/\mathcal{P}_T$, and the peak scale $k_{\text{peak}}$
are summarized in Table \ref{table1},
and the scalar power spectra for these models are shown in Fig. \ref{pr}.
We also show the peak values of the power spectra in Table \ref{table2}.
From these results, we see that the results $n_s\approx 0.968$ and $r\approx 0.04$
at the pivotal scale $k_*=0.05$ Mpc$^{-1}$ are consistent with the observational constraints \cite{Akrami:2018odb}
\begin{equation}
\label{cmb:con}
n_s = 0.9649\pm 0.0042  ~(68\% \text{CL}) ,\quad
r_{0.05} < 0.06 ~(95\% \text{CL}).
\end{equation}
As discussed above and in Ref. \cite{Yi:2020kmq}, the parameter $q$ controls the shape of the peak and the parameter $\phi_p$ determines the peak scales.
By choosing different values of $\phi_p$ and $q$, we can produce either
sharp peaks or broad peaks with different peak scales.

\begin{table*}
\caption{The model parameters and the numerical results.}
\label{table1}
\begin{tabular}{ccccccccc}
\hline
		Model &$\Lambda^4/10^{-9}$ &$h/10^{10}$ &$w/10^{-11}$ &$\phi_p$ &$N$&$n_s$&$r$&$k_{\text{peak}}/\text{Mpc}^{-1}$\\
		\hline
 		N1  &$ 2.69$~&$ 1.55$ & 1.08 & $15.54$~&$61.98$~&$0.963$~&$0.047$~&$3.22\times 10^{12}$\\
        N2  &$ 2.73$~&$ 1.33$ & 1.14 & $15.05$&$61.21$&$0.964$&$0.046$&$8.90\times 10^{8}$\\
        N3   &$ 2.74$~&$ 1.432$ & 1.01 & $14.78$&$61.05$&$0.968$&$0.045$&$ 9.38\times 10^{5}$\\
        WN1   &$ 2.79$~&$ 21.2$ & 2.27 & $16.01$&$70.82$&$0.962$&$0.047$&$ 4.41\times 10^{12}$\\
        WN2   &$ 2.77$~&$ 1.55$ &15.6 & $15.25$&$66.32$&$0.962$&$0.047$&$8.22\times 10^{8}$\\
        WN3    &$ 2.87$~&$ 21.06$ & 1.56 & $14.88$&$65.02$&$0.962$&$0.047$&$ 3.95\times 10^{5}$\\
\hline
\end{tabular}
\end{table*}

\begin{figure}[htbp]
$\begin{array}{cc}
\includegraphics[width=0.46\textwidth]{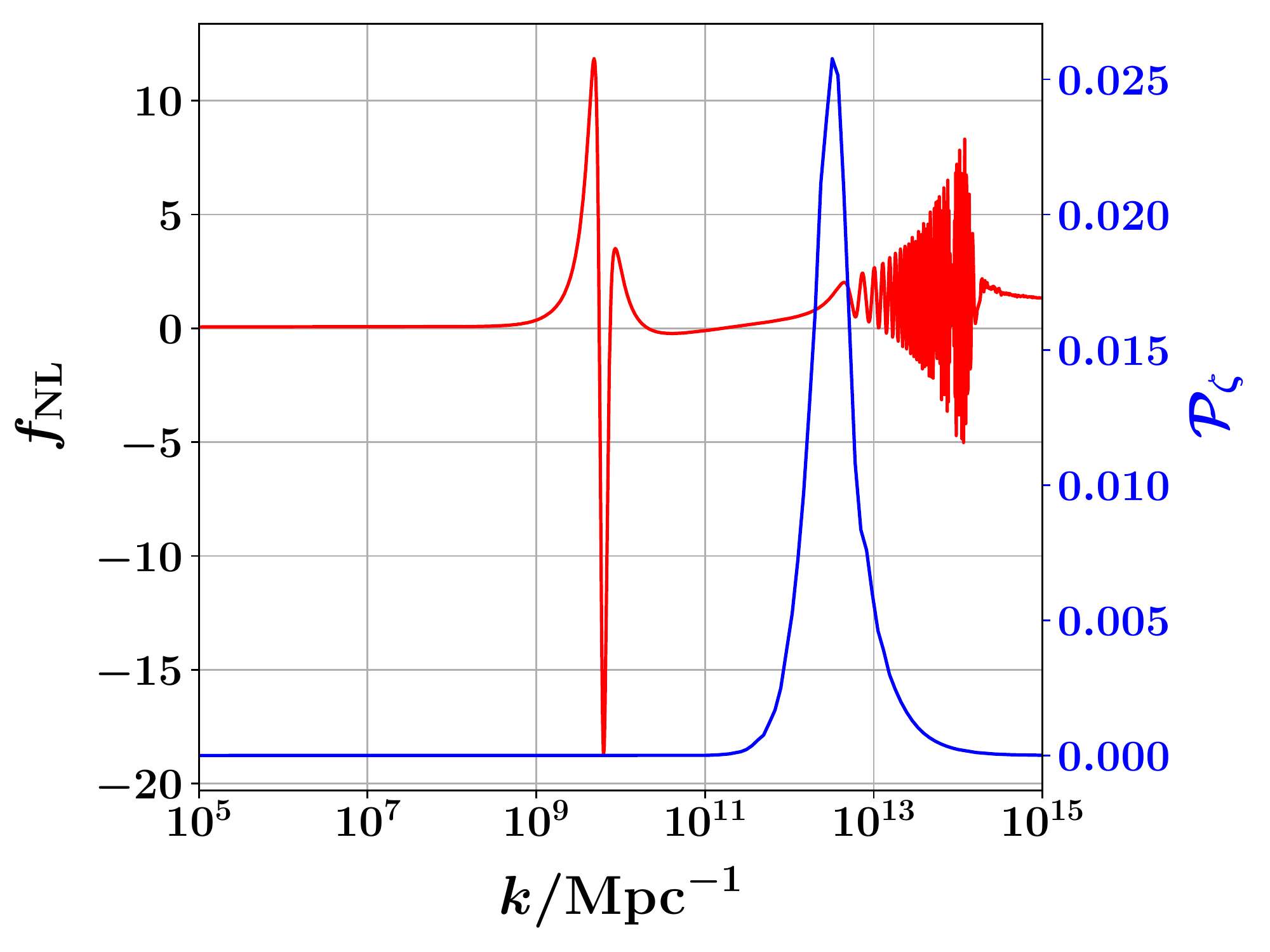}&
\includegraphics[width=0.46\textwidth]{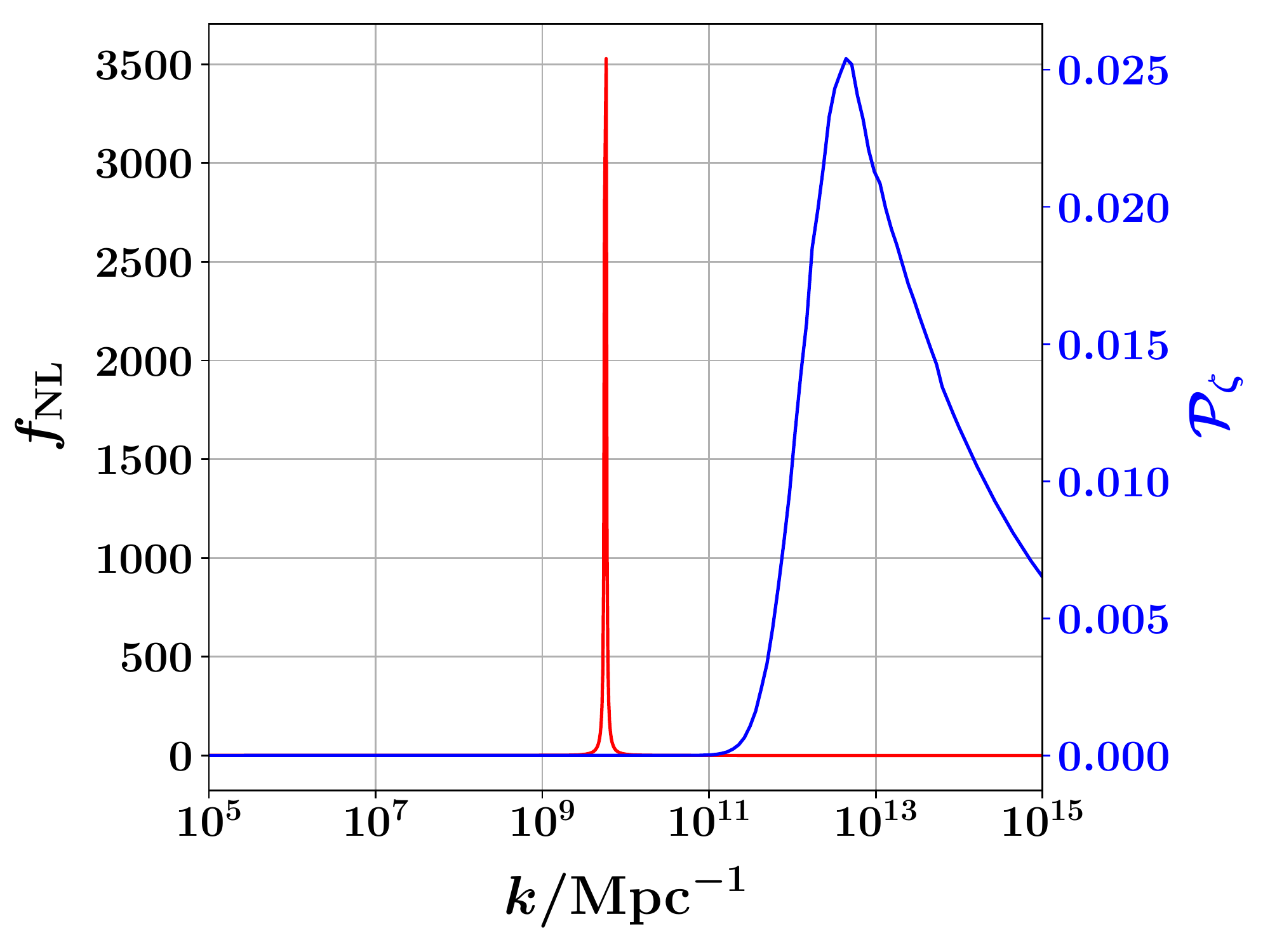}
\end{array}$
\caption{The results of the non-Gaussianity parameters $f_{\text{NL}}$ (red lines) in the equilateral limit along with the primordial scalar power spectrum $\mathcal{P}_{\zeta}$  (blue lines) for the models N1 and WN1.
  The left panel shows the results for the model N1 and the right panel shows the results for the model WN1.}\label{neq}
\end{figure}

\begin{figure}[htbp]
  \centering
  \includegraphics[width=0.7\textwidth]{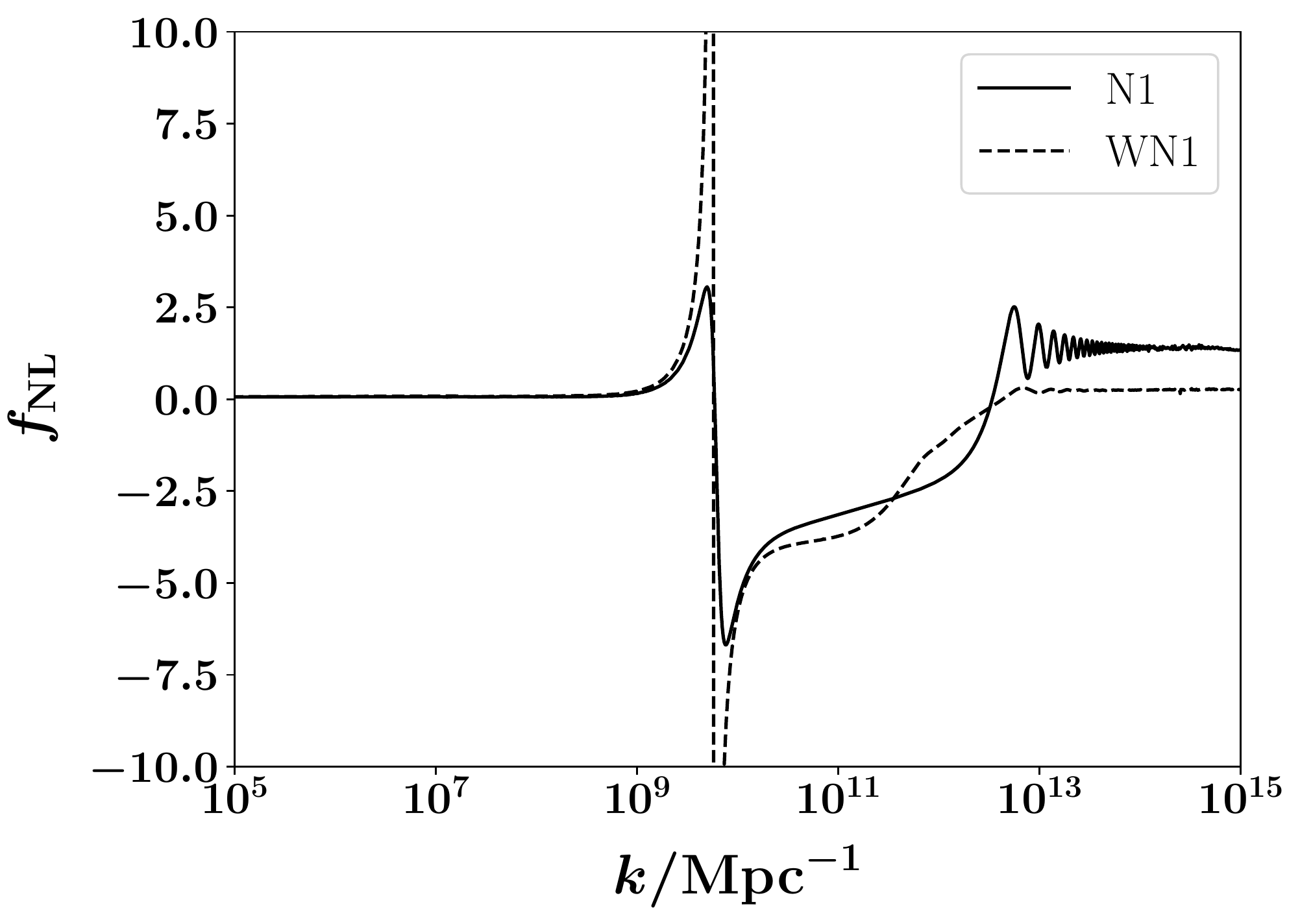}
  \caption{The results of the non-Gaussianity parameters $f_{\text{NL}}$ in the squeezed limit for the models N1 (solid line) and WN1 (dashed line).}\label{nsq}
\end{figure}

We also calculate the three-point correlation function numerically to get the bispectrum $B_{\zeta}$ \cite{Byrnes:2010ft,Ade:2015ava,Zhang:2020uek},
\begin{equation}\label{Bi}
\left\langle\hat{\zeta}_{\bm{k}_{1}}\hat{\zeta}_{\bm{k}_{2}}\hat{\zeta}_{\bm{k}_{3}}\right\rangle=(2 \pi)^{3} \delta^{3}\left(\bm{k}_{1}+\bm{k}_{2}+\bm{k}_{3}\right) B_{\zeta}\left(k_{1}, k_{2}, k_{3}\right),
\end{equation}
where $\hat{\zeta}_{\bm{k}}$ is the corresponding quantum operator of the curvature perturbation $\zeta_k$.
The non-Gaussianity parameter $f_\text{NL}$ is \cite{Byrnes:2010ft,Creminelli:2006rz}
\begin{equation}\label{Fnl}
f_{\text{NL}} (k_1,k_2,k_3)=\frac{5}{6}\frac{B_{\zeta}(k_1,k_2,k_3)}{P_{\zeta}(k_1)
P_{\zeta}(k_2)+P_{\zeta}(k_2)P_{\zeta}(k_3)+P_{\zeta}(k_3)P_{\zeta}(k_1)},
\end{equation}
where $P_{\zeta}(k)=|\zeta_k|^2=2\pi^2\mathcal{P}_\zeta/k^3$. The non-Gaussianity parameter $f_\text{NL}$ in the equilateral and squeezed limits are shown in Figs. \ref{neq} and \ref{nsq}, respectively.

\subsection{Primordial black hole dark matter}

During radiation dominated era, PBHs may form by gravitational collapse from the enhanced primordial curvature perturbations at small scales when they reenter the horizon.
Ignoring the mass accretion and evaporation in the Press-Schechter formalism,
the fractional energy density of PBHs
at the formation is \cite{Young:2014ana,Ozsoy:2018flq,Tada:2019amh}
\begin{equation}
\label{eq:beta1}
\beta(M) \approx \sqrt{\frac{2}{\pi}}\frac{\sigma(M)}{\delta_c}
\exp\left(-\frac{\delta_c^2}{2\sigma^2(M)}\right),
\end{equation}
where the critical density perturbation for the PBH formation is $\delta_c=0.4$ \cite{Tada:2019amh,Harada:2013epa,Escriva:2019phb,Yoo:2020lmg} and the mass variance $\sigma(M)$ associated with the PBH mass is
\begin{equation}\label{simga1}
\sigma(R)^2=\int_{0}^{\infty}W^2(k R)\frac{\mathcal{P}_\delta(k)}{k}dk,
\end{equation}
$W(kR)$ is the window function to smooth the density contrast,
the smoothing scale $R=(aH)^{-1}$,
and the power spectrum $\mathcal{P}_\delta$ of the matter perturbation is related with the primordial curvature as
\begin{equation}
\label{rel:pp}
\mathcal{P}_\delta(k)=\frac{16}{81}\left(\frac{k}{aH}\right)^4 \mathcal{P}_{\zeta}(k).
\end{equation}
Although there exist different window functions \cite{Ando:2018qdb},
we take  the Gaussian  window function
\begin{equation}\label{Gauus:window}
 W(k R)=\exp\left(-k^2R^2/2\right).
 \end{equation}
Substituting the Gaussian window function \eqref{Gauus:window} and Eq. \eqref{rel:pp} into Eq. \eqref{simga1}, the mass variance becomes
\begin{equation}\label{phb:variance}
\sigma^2(R)=\frac{4(1+w)^2}{(5+3w)^2}\int_{0}^{\infty}x^3 \exp\left(-x^2\right) \mathcal{P}_{\zeta}(x/R)dx,
\end{equation}
where $x=k R$.
Note that the main contribution to the integral \eqref{phb:variance} is from the scale $x\sim 1$
and the result is not affected much by other scales,
so to a good approximation,
the primordial curvature  power spectrum can be treated as scale invariant even though it may change rapidly \cite{Sato-Polito:2019hws}.
With this assumption, the mass variance \eqref{phb:variance} becomes
\begin{equation}\label{phb:variance2}
\sigma^2\approx\frac{8}{81} \mathcal{P}_{\zeta}(1/R).
\end{equation}
Substituting Eq. \eqref{phb:variance2} into the definition  \eqref{eq:beta1}, we obtain
\begin{equation}
\label{eq:beta}
\beta(M) \approx \sqrt{\frac{2}{\pi}}\frac{\sqrt{P_{\zeta}}}{\mu_c}
\exp\left(-\frac{\mu_c^2}{2P_{\zeta}}\right),
\end{equation}
where $\mu_c=9\delta_c/2\sqrt{2}$.
The current fractional energy density of PBHs with mass $M$ to DM \cite{Carr:2016drx,Gong:2017qlj}
\begin{equation}
\label{fpbheq1}
\begin{split}
Y_{\text{PBH}}(M)=&\frac{\beta(M)}{3.94\times10^{-9}}\left(\frac{\gamma}{0.2}\right)^{1/2}
\left(\frac{g_*}{10.75}\right)^{-1/4}\\
&\times \left(\frac{0.12}{\Omega_{\text{DM}}h^2}\right)
\left(\frac{M}{M_\odot}\right)^{-1/2},
\end{split}
\end{equation}
where $M_{\odot}$ is the solar mass, $\gamma= 0.2$ \cite{Carr:1975qj}, the current
energy density parameter of DM is taken to be $\Omega_{\text{DM}}h^2=0.12$ \cite{Aghanim:2018eyx},
the effective degrees of freedom $g_*=107.5$ for $T>300$ GeV
and $g_*=10.75$ for $0.5\ \text{MeV}<T<300\ \text{GeV}$.
The peak mass scale $M$ for PBHs and the peak scale $k_{\text{peak}}$
is estimated to be \cite{Gong:2017qlj}
\begin{equation}
\label{mkeq1}
M(k)=3.68\left(\frac{\gamma}{0.2}\right)\left(\frac{g_*}{10.75}\right)^{-1/6}
\left(\frac{k_\text{peak}}{10^6\ \text{Mpc}^{-1}}\right)^{-2} M_{\odot}.
\end{equation}

Note that there are subtleties for the simple relationship
between $\beta(M)$ and $\mathcal{P}_\zeta$.
When we consider the non linearities between the Gaussian
curvature perturbation and the density contrast and the non-linear effects arising at horizon crossing,
the value of $\delta_c$ may become larger \cite{Musco:2020jjb}.
On the other hand, the abundance of PBHs also depends on the shape and non-Gaussianity of $\mathcal{P}_{\zeta}$ and non-linear statistics \cite{Atal:2018neu,Germani:2018jgr,Germani:2019zez,Escriva:2021pmf}.
If we consider non-Gaussianities of the primordial power spectrum,
then the PBH abundance at the peak is \cite{Saito:2008em}
\begin{equation}
\label{beta}
\beta(M_\text{peak})\approx \frac{1}{\sqrt{2\pi}}\int_{\tilde{\zeta}_{th}}[(\tilde{\zeta}^2-1)-(\tilde{\zeta}^5-8\tilde{\zeta}^3+9\tilde{\zeta})
\mathcal{J}_\mathrm{peak}]e^{-\tilde{\zeta}^2/2},
\end{equation}
where $\tilde{\zeta}=\zeta/\sqrt{\mathcal{P}_\zeta(k_\text{peak})}$ and
\begin{equation}
\label{jeq2}
\mathcal{J}_\mathrm{peak}=\frac{3}{20\pi}f_\mathrm{NL}(k_\mathrm{peak},k_\mathrm{peak},k_\mathrm{peak})\sqrt{\mathcal{P}_\zeta(k_\mathrm{peak})}.
\end{equation}
For a good estimation, we can use the parameter $\mathcal{J}_\mathrm{peak}$
to characterize the effect of the non-Gaussianities.
If $\mathcal{J}_\mathrm{peak}\ll 1$, then we can ignore the effect of non-Gaussianities on the PBH abundance.
From Figs. \ref{neq} and \ref{nsq}, we see that non-Gaussianities around the peak scale are small and henceforth $\mathcal{J}_\mathrm{peak}\ll 1$, so the effect of non-Gaussianities on the PBH abundance is negligible.

\begin{figure}[htbp]
  \centering
  \includegraphics[width=0.7\textwidth]{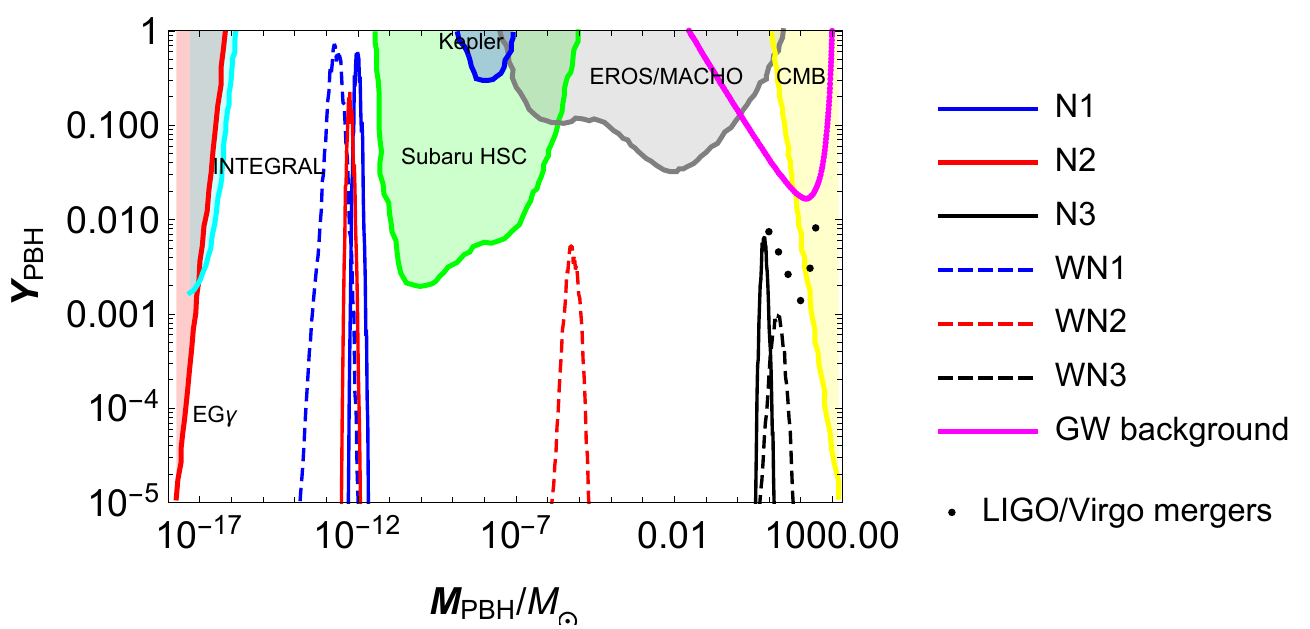}
  \caption{The results for PBH abundances.
  The parameters for the models are shown in Table \ref{table1},
  the peak abundance and the peak mass of PBHs are shown in Table \ref{table2}.
The shaded regions show the observational constraints on the PBH abundance:
the yellow region from accretion constraints by CMB \cite{Ali-Haimoud:2016mbv,Poulin:2017bwe},
the red region from extragalactic gamma-rays by PBH evaporation \cite{Carr:2009jm} (EG$\gamma$), the cyan region from galactic center 511 keV gamma-ray line (INTEGRAL) \cite{Laha:2019ssq,Dasgupta:2019cae},
the green region from microlensing events with Subaru HSC \cite{Niikura:2017zjd},
the blue region from the Kepler satellite \cite{Griest:2013esa},
the gray region from the EROS/MACHO \cite{Tisserand:2006zx}.
The magenta solid line shows the constraints on the stochastic gravitational wave background by LIGO/Virgo \cite{LIGOScientific:2018mvr,Raidal:2017mfl} and the black dots shows the  limits from the LIGO/Virgo merger rate \cite{LIGOScientific:2018mvr,Ali-Haimoud:2017rtz}.}\label{fpbh}
\end{figure}

Substituting the numerical results of the power spectra obtained in the previous subsection into Eqs. \eqref{eq:beta}, \eqref{fpbheq1} and \eqref{mkeq1},
we obtain the abundance and the peak mass of PBHs and the results are shown in Fig. \ref{fpbh} and Table \ref{table2}.
In Fig. \ref{fpbh}, the constraint on the abundance of PBHs from white dwarf explosion \cite{Graham:2015apa} is not included
because it is not robust \cite{Montero-Camacho:2019jte}.
From Fig. \ref{fpbh} and Table \ref{table2}, we see that
different peak scales in the scalar power spectrum correspond to different masses of PBHs.
In particular, the models generate PBHs with masses around $10^{-13}\ M_{\odot}$,
$10^{-6}\ M_{\odot}$ and $10\ M_{\odot}$, respectively.
The stellar mass PBHs may explain BHs observed by LIGO/Virgo collaboration.
The peak abundance of PBHs with the mass scale $10^{-13}\ M_{\odot}$  is $Y_\text{PBH}^\text{peak}\approx 1$, so they can
make up almost all the DM.
PBH DM with the mass around $\mathcal{O}(1)M_{\oplus}$ may explain the planet 9.

\begin{table}[htp]
\caption{The results for the peak value of the primordial scalar power spectra, the peak mass and peak abundance of PBH and the peak frequency of SIGWs.
The parameters for the models are shown in Table \ref{table1}.}
	\renewcommand\tabcolsep{4.0pt}
	\begin{tabular}{ccccc}
		\hline
		Model \quad  &$\mathcal{P}_{\zeta(\text{peak})}$& $M_\text{peak}/M_\odot$&$Y_\text{PBH}^\text{peak}$& $f_c/\text{Hz}$\\
		\hline
 		N1 \quad   &$0.0258$&$3.54\times10^{-13}$&$0.954$&$5.83\times 10^{-3}$\\
        N2 \quad   &$0.0258$&$4.65\times10^{-6}$&$2.88\times 10^{-4}$&$1.61\times 10^{-6}$\\
        N3 \quad   &$0.0357$&$4.19$&$2.11\times 10^{-3}$&$2.35\times 10^{-8}$\\
        WN1 \quad   &$0.0254$&$1.90\times10^{-13}$&$0.829$&$7.97\times 10^{-3}$\\
        WN2 \quad   &$0.0269$&$4.41\times 10^{-6}$&$1.02\times10^{-3}$&$1.57\times10^{-6}$\\
        WN3 \quad   &$0.0369$&$23.56$&$1.81\times10^{-3}$&$3.56\times 10^{-8}$\\
		\hline
	\end{tabular}
	
\label{table2}
\end{table}

\subsection{Scalar induced secondary gravitational waves}

\begin{figure}[htbp]
  \centering
  \includegraphics[width=0.7\textwidth]{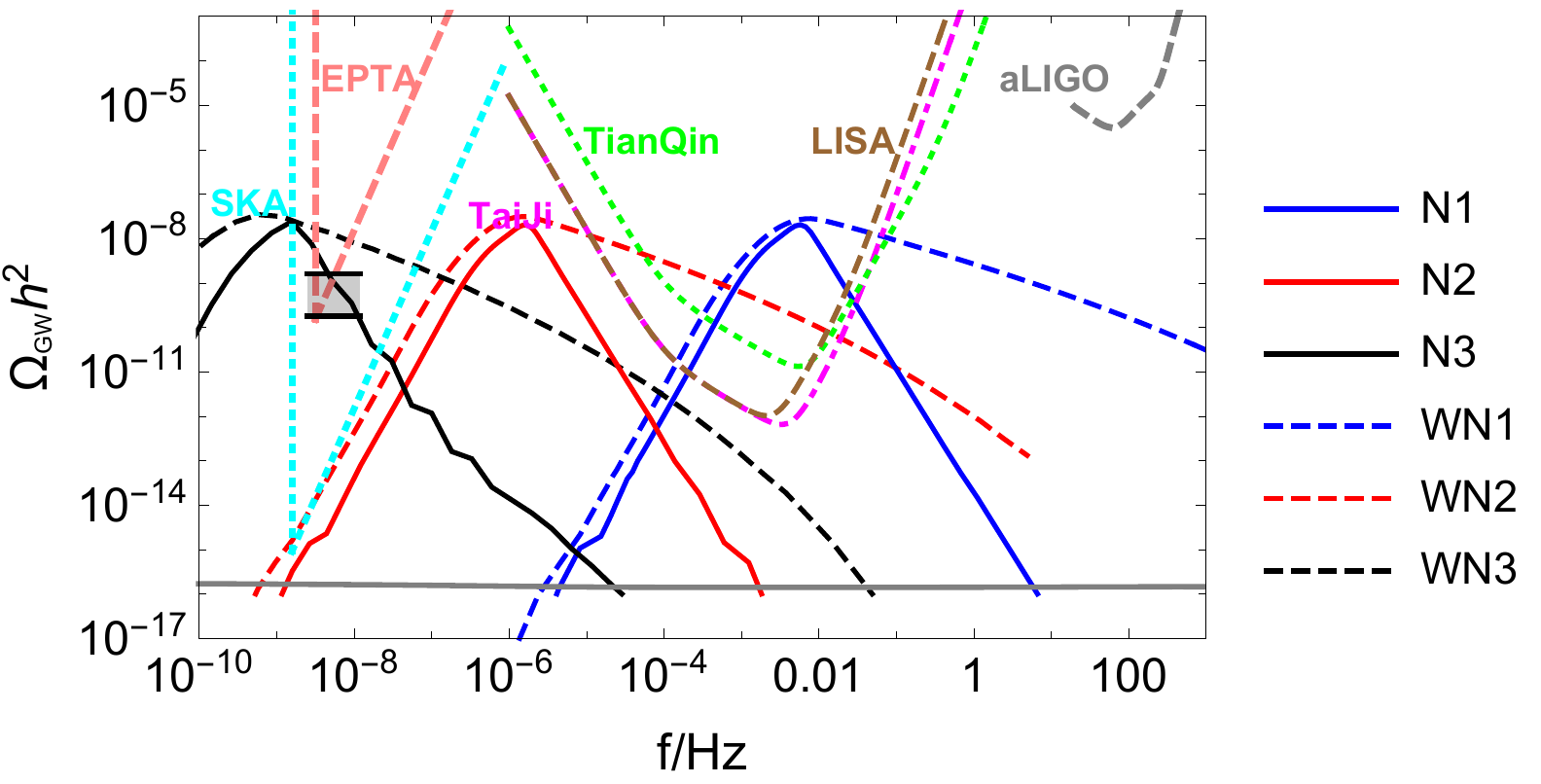}
  \caption{The energy densities of SIGWs.
  The parameters for the models are shown in Table \ref{table1}
  and the peak frequencies are shown in Table \ref{table2}.
The dashed pink curve denotes the EPTA limit \cite{Ferdman:2010xq,Hobbs:2009yy,McLaughlin:2013ira,Hobbs:2013aka} ,
the dotted cyan curve denotes the SKA limit \cite{Moore:2014lga},
the shaded region is the observational result from NANOGrav 12.5-year data \cite{Arzoumanian:2020vkk},
the dashed green curve in the middle denotes the TianQin limit \cite{Luo:2015ght},
the dot-dashed magenta curve shows the TaiJi limit \cite{Hu:2017mde},
the dashed brown curve shows the LISA limit \cite{Audley:2017drz},
and the dashed gray curve denotes the aLIGO limit \cite{Harry:2010zz,TheLIGOScientific:2014jea}.
For comparison, the energy density of primordial GWs from the model N1 is also shown with the gray solid line.
}\label{gw}
\end{figure}

For the perturbation to the second order,
the first order and second order perturbations are mixed and the first order scalar perturbations become the source of the second order tensor perturbations.
Therefore, the large primordial scalar perturbations at small scales
induce secondary GWs
associated with the production of PBHs when they reenter the horizon.
The equation for the
Fourier components of the second order tensor perturbations $h_{\bm{k}}$ is
\cite{Ananda:2006af,Baumann:2007zm}
\begin{equation}
\label{eq:hk}
h''_{\bm{k}}+2\mathcal{H}h'_{\bm{k}}+k^2h_{\bm{k}}=4S_{\bm{k}},
\end{equation}
where the source from the first order scalar perturbations is
\begin{equation}
\label{hksource}
\begin{split}
S_{\bm{k}}=\int \frac{d^3\tilde{k}}{(2\pi)^{3/2}}e_{ij}(\bm{k})\tilde{k}^i\tilde{k}^j
\left[2\Phi_{\tilde{\bm{k}}}\Phi_{\bm{k}-\tilde{\bm{k}}}
+\frac{4}{3(1+w)\mathcal{H}^2} \right.\\
\left. \times\left(\Phi'_{\tilde{\bm{k}}}+\mathcal{H}\Phi_{\tilde{\bm{k}}}\right)
\left(\Phi'_{\bm{k}-\tilde{\bm{k}}}+\mathcal{H}\Phi_{\bm{k}-\tilde{\bm{k}}}\right)\right],
\end{split}
\end{equation}
$\mathcal{H}= aH$, $w=p/\rho=1/3$ during radiation domination, $e_{ij}(\bm{k})$ is the polarization tensor, $\Phi$ is the gauge invariant Bardeen potential.
The energy density of SIGWs
in the radiation domination is \cite{Inomata:2016rbd,Kohri:2018awv,Espinosa:2018eve}
\begin{equation}
\label{gwres1}
\begin{split}
\Omega_{\mathrm{GW}}(k,\eta)=&\frac{1}{6}\left(\frac{k}{aH}\right)^2
\int_{0}^{\infty}dv\int_{|1-v|}^{1+v}du\left\{ \right.\\
&\left[\frac{4v^2-(1-u^2+v^2)^2}{4uv}\right]^2\\
&\left. \times \overline{I_{\text{RD}}^{2}(u, v, x\to \infty)} P_{\zeta}(kv)P_{\zeta}(ku)\right\},
\end{split}
\end{equation}
where the kernel function $\overline{I_{\text{RD}}^{2}}$ is \cite{Espinosa:2018eve,Lu:2019sti}
\begin{equation}
\label{irdeq}
\begin{split}
    \overline{I^2_{\text{RD}}(u,v,x\rightarrow\infty)}=&\frac{1}{2x^2}\left[\left(\frac{3\pi(u^2+v^2-3)^2\Theta(u+v-\sqrt3)}{4u^3v^3}+\frac{T_c(u,v,1)}{9}\right)^2\right.\\
    &\qquad \left.+\left(\frac{\tilde{T}_s(u,v,1)}{9}\right)^2\right],
\end{split}
\end{equation}
$T_c(u,v,x)$ and $\tilde{T}_s(u,v,x)$ are given in Ref. \cite{Lu:2019sti}.
The current energy density of SIGWs is
\begin{equation}\label{gwres2}
\Omega_{GW}(k,\eta_0)=\Omega_{GW}(k,\eta)\frac{\Omega_{r}(\eta_0)}
{\Omega_{r}(\eta)},
\end{equation}
where $\Omega_r$ is the fraction energy density of radiation.
Plugging the power spectra in Fig. \ref{pr} into Eqs. \eqref{gwres1} and \eqref{gwres2}
and using Eq. \eqref{irdeq} we obtain current energy densities of SIGWs
and the results are shown in Fig. \ref{gw}.
The peak frequencies  are shown in Table \ref{table2}.
From Table \ref{table2}, we see that the peak frequencies of the SIGWs are around mHz, $\mu$Hz and nHz respectively.
For the models WN1, WN2 and WN3, $\Omega_{\text{GW}}$ has a broad shape which spans a wide frequency bands because the enhanced power spectrum has a broad peak.
The broad shape for the model WN3 which produces the stellar mass PBHs with the abundance in the order of $10^{-3}$ leads to its
exclusion by the EPTA data \cite{Ferdman:2010xq,Hobbs:2009yy,McLaughlin:2013ira,Hobbs:2013aka},
but the broad shape for the model WN2 which produces
the earth-mass PBHs to explain the planet 9 makes the model testable by LISA and Taiji.
The models N1 and WN1 can explain DM in terms of PBHs and they can be tested by LISA/Taiji/TianQin.
The model N3 may explain the possible stochastic GW background detected in the North American Nanohertz Observatory for Gravitational Waves (NANOGrav) 12.5-year data \cite{Arzoumanian:2020vkk}.
If the abundance of PBHs produced in the model WN3 is in the order of $10^{-15}$,
then the model WN3 can also explain the possible stochastic GW background detected in the NANOGrav 12.5-year data.
For comparison, we also show the energy density of the primordial GWs from the model N1 in Fig. \ref{gw}. The energy density of the primordial GWs is in the order of $10^{-16}$ and
the peak values of the energy densities of the SIGWs generated in our models are in the order of  $10^{-8}$.

\section{Conclusion}
\label{sec3}

The amplitude of primordial curvature perturbations at small scales can become large by the enhancement mechanism with
a peak function in the non-canonical kinetic term \cite{Lin:2020goi,Yi:2020kmq,Yi:2020cut}.
We apply the enhancement mechanism to natural inflation to produce abundant PBHs and observable SIGWs.
The power spectrum at large scales is consistent with observational constraint
and the power spectrum at small scales is enhanced to the order of 0.01.
Either sharp peak or broad peak is possible with different peak shapes for the peak function by choosing different values of $q$.
By adjusting the peak position $\phi_p$ in the peak function, the power spectrum
is enhanced at different scales,
henceforth associated with the generation of SIGWs with different peak frequencies, PBHs with different masses are produced.
We choose three different values of $\phi_p$ to get enhanced power spectrum at
$10^{12}$ Mpc$^{-1}$, $10^{8}$ Mpc$^{-1}$ and $10^{5}$ Mpc$^{-1}$, respectively.
The enhanced curvature perturbations produce PBH DM with peak masses around
$10^{-13}\ M_{\odot}$, the Earth's mass and the stellar mass,
and SIGWs with peak frequencies around mHz, $\mu$Hz and nHz.
The stellar mass PBHs may explain BHs observed by LIGO/Virgo collaboration,
and the earth-mass PBHs may explain the planet 9.
The PBHs with the mass scale $10^{-13}\ M_{\odot}$ can
make up almost all the DM.
The peak energy densities of SIGWs for the models discussed in this paper are around $10^{-8}$ while the energy density of primordial GWs is around $10^{-16}$.
The SIGWs with the peak frequency around nHz is testable by PTA observations,
and SIGWs with the peak frequency around mHz is testable by space based GW observatory.
In particular, the SIGWs produced in the model N3 can explain the stochastic GW background observed by NANOGrav 12.5-year experiment.
The broad shape for the model WN3 which produces the stellar mass PBHs with the abundance in the order of $10^{-3}$  leads to its
exclusion by the EPTA data
and the broad shape for the model WN2 makes the model testable by LISA and Taiji.
If the abundance of PBHs produced in the model WN3 is in the order of $10^{-15}$,
then the model WN3 can also explain the possible stochastic GW background detected in the NANOGrav 12.5-year data.
These results are similar to those obtained in \cite{Lin:2020goi,Yi:2020kmq,Yi:2020cut,Gao:2021vxb}.
The detailed shape of the primordial power spectrum and non-Gaussianity in different models are not exactly the same, more accurate measurements may be able to distinguish different models.

In conclusion, the enhancement mechanism with
a peak function in the non-canonical kinetic term
works for natural inflation.

\section*{CRediT authorship contribution statement}

{\bf Q. Gao:} Formal analysis, Methodology, Software, Writing - original draft,
Writing - review \& editing.
{\bf Y. Gong:} Conceptualization, Formal analysis, Methodology, Software, Supervision, Writing - original draft, Writing - review \& editing.
{\bf Z. Yi:} Formal analysis, Software, Writing - review \& editing.

\section*{Declaration of competing interest}

The authors declare that they have no known competing financial interests or personal relationships that could have appeared to influence the work reported in this paper.

\section*{Acknowledgements}
\label{acknowledgements}

This research is supported in part by the National Key Research \& Development Program of China under Grant No. 2020YFC2201504,
the Venture \& Innovation Support Program for Chongqing Overseas Returnees under Grant No. CX2020083,
the National Natural Science
Foundation of China under Grant No. 11875136,
the Major Program of the National Natural Science Foundation of China under Grant No. 11690021,
and MOE Key Laboratory of TianQin Project, Sun Yat-sen University.
Z. Yi acknowledges support by China Postdoctoral Science Foundation Funded Project under Grant No. 2019M660514.





\end{document}